\documentclass{IOS-Book-Article}

\usepackage{mathptmx}
\usepackage{soul}\setuldepth{article}
\usepackage{enumitem}
\usepackage{latexsym}
\usepackage{amssymb}
\usepackage{amsmath}
\usepackage{amsthm}
\usepackage{booktabs}
\usepackage{graphicx}
\usepackage{color}
\usepackage{tcolorbox}

\def\hb{\hbox to 11.5 cm{}}

\begin{document}

\pagestyle{headings}
\def\thepage{}
\begin{frontmatter}              

\title{Zero-Shot Topic Classification \\ 
of Column Headers: \\
Leveraging LLMs for Metadata Enrichment}

\markboth{}{April 2024\hb}

\author{\fnms{Margherita}~\snm{Martorana}\orcid{0000-0001-8004-0464}%
\thanks{Corresponding Author: Margherita Martorana, m.martorana@vu.nl}},
\author{\fnms{Tobias} \snm{Kuhn}\orcid{0000-0002-1267-0234}}, 
\author{\fnms{Lise} \snm{Stork}\orcid{0000-0002-2146-4803}},
\author{\fnms{Jacco}~\snm{van Ossenbruggen}\orcid{0000-0002-7748-4715}}

\runningauthor{M. Martorana et al.}
\address{Department of Computer Science, Vrije Universiteit Amsterdam, De Boelelaan 1105, Amsterdam, The Netherlands}

\begin{abstract}
Traditional dataset retrieval systems rely on metadata for indexing, rather than on the underlying data values. However, high-quality metadata creation and enrichment often require manual annotations, which is a labour-intensive and challenging process to automate. In this study, we propose a method to support metadata enrichment using topic annotations generated by three Large Language Models (LLMs): ChatGPT-3.5, GoogleBard, and GoogleGemini. Our analysis focuses on classifying column headers based on domain-specific topics from the Consortium of European Social Science Data Archives (CESSDA), a Linked Data controlled vocabulary. Our approach operates in a zero-shot setting, integrating the controlled topic vocabulary directly within the input prompt. This integration serves as a Large Context Windows approach, with the aim of improving the results of the topic classification task.

We evaluated the performance of the LLMs in terms of internal consistency, inter-machine alignment, and agreement with human classification. Additionally, we investigate the impact of contextual information (i.e., dataset description) on the classification outcomes. Our findings suggest that ChatGPT and GoogleGemini outperform GoogleBard in terms of internal consistency as well as LLM-human-agreement. Interestingly, we found that contextual information had no significant impact on LLM performance. 

This work proposes a novel approach that leverages LLMs for topic classification of column headers using a controlled vocabulary, presenting a practical application of LLMs and Large Context Windows within the Semantic Web domain. This approach has the potential to facilitate automated metadata enrichment, thereby enhancing dataset retrieval and the Findability, Accessibility, Interoperability, and Reusability (FAIR) of research data on the Web.
\end{abstract}

\end{frontmatter}
\markboth{April 2022\hb}{April 2022\hb}


\section{Introduction}
Traditional dataset retrieval systems index on metadata information rather than on the underlying data values. Despite the critical role of high-quality metadata for data retrieval, many datasets still lack informative metrics and annotations to facilitate their discovery \cite{vlachidis2021semantic}. Creating and enriching metadata with high-quality information and annotations is a labour intensive and challenging process to automate, and it often relies on knowledge from domain experts. Enriching metadata with column-level information, such as with the topic described by each column, poses particular difficulty due to sparse contextual information and reliance on domain-specific codebooks, often not available in digital structured format. Moreover, column-level information becomes even more critical in the context of restricted access datasets, where users cannot directly investigate the underlying data due to confidentiality issues. In such cases, the availability of high-quality metadata with detailed column-level information becomes even more a primary need to assess the relevance and suitability of the datasets retrieved. 

The FAIR Guiding Principles \cite{10.1038/sdata.2016.18} emphasise the importance of high-quality metadata to facilitate the Findability, Accessibility, Interoperability, and Reusability (FAIR) of data on the Web. Several studies have found that by applying the FAIR Principles, we can not only improve data management and stewardship \cite{boeckhout2018fair,mons2018data}, but also facilitate data transparency, reproducibility, discovery and reuse \cite{wilkinson2017interoperability} and resource citation \cite{lamprecht2020towards}. Recently, there has also been a concrete effort to incorporate column-level information into metadata schemas, recognising its essential role in facilitating the discovery and reuse of datasets \cite{martorana2023advancing}. However, the sparsity and fragmentation of information that can occur in the context of restricted access data leads to challenges in applying traditional topic classification techniques. 

The rise of advanced Large Language Models (LLMs) has presented several opportunities and challenges in automating data annotation and metadata creation \cite{tan2024large}. Studies have shown some preliminary results regarding the advantages and disadvantages of various LLMs and overall performance variations \cite{singh2023chat,rane2024gemini}. However, it is still not clear how different LLMs perform in topic classification tasks, particularly when dealing with short texts such as column headers, and in the context of restricted access data. 

\subsection{Use Case}
To illustrate the motivation behind this research, consider the following scenario. A socioeconomic researcher is investigating the relationship between income inequalities and proximity to higher education institutions, and may need to use multiple datasets or fragments of datasets. However, given that socioeconomic data are likely to be confidential, direct examination may not be possible. In this context, the availability of high-quality metadata with rich column-level information is crucial to discover and explore common attributes across multiple datasets. For example, column metadata could be enriched with the CESSDA topic controlled vocabulary, which include terms related to both the topics of \textit{`income inequality'} and \textit{`education'}. Without rich metadata, the researcher would face significant challenges in identifying relevant datasets. 

\subsection{Research questions and Contributions}
In this work we address the challenges of automated metadata enrichment in the context of restricted access data, by investigating how we can leverage Large Language Models in a zero-shot setting and by also following a Large Context Window approach. Specifically, we explore how LLMs can perform the column header topic classification task by using a controlled vocabulary of topics. In this approach, no fine-tuning of the models is necessary, as the controlled vocabulary is provided directly as part of the input. The controlled vocabulary of topics is used for the column header classification task, leveraging a Large Context Window approach and a zero-shot setting. It is important to note that because our research focuses on the domain of restricted access data, we will only use the column headers during our analysis and not the underlying data. Additionally, we will investigate whether incorporating contextual information about the datasets - i.e. the dataset descriptions provided by the publisher - results in any differences in the classification task. To guide our investigation, we formulate the following research questions:

\begin{enumerate}
    \item What is the consistency of the LLMs in the topic classification task of column headers from a controlled vocabulary?
    \item What are the difference in the topic classification task of column headers between LLMs and humans?
    \item Do hierarchical and contextual information have any effect in the classification task of column headers?
\end{enumerate}

Our work contributes to the current knowledge on the applications of LLMs by assessing the performance of three LLMs (GPT-3.5, GoogleBard, and GoogleGemini) in the topic classification task of column headers with a controlled vocabulary, and comparing it with human-made classifications. To the best of our knowledge, this is the first work to investigate the performance of various LLMs in this specific task and under these settings. 


\section{Related Work}

\subsection{Semantic Metadata Enrichment}
Semantic metadata enrichment refers to the process of enhancing metadata with additional meanings and contexts to improve both human- and machine-readability. This generally involves the incorporation of semantic annotations derived from ontologies~\cite{lombardo2013ontologies,bernasconi2018ontology}, thesauri~\cite{koutsomitropoulos2018learning,koutsomitropoulos2019semantic}, or specialised controlled vocabularies~\cite{lisena2018controlled,gil2017controlled,afzal2008towards}, which enrich the content and connections with external resources. Unlike simpler annotations, semantic enrichment provides a deeper layer of context and structure. Previous research shows that the application of semantic metadata enrichment and the use of controlled vocabularies helps in the FAIRification of data on the web and fosters cross-disciplinary cooperation between research entities and institutions \cite{vlachidis2021semantic,sasse2022semantic}. Furthermore, it has previously been shown how the semantic annotation of metadata, such as column-level metadata \cite{dugas2016odmedit,martorana2023advancing,magagna2021adopt}, can improve the Findability, Accessibility, Interoperability, and Reusability of research data \cite{jonquet2023common}, a crucial aspect for reusing data with restricted access (i.e., medical records and microdata \cite{razick2014egenvar}), which typically lacks detailed information due to its sensitive and confidential nature. Semantic annotations at the column level can facilitate the discovery of such data by adding more context while adhering to privacy preservation standards \cite{dugas2016odmedit}. 

\subsection{Topic Classification with LLMs}
Large Language Models (LLMs) have revolutionised the field of Natural Language Processing (NLP), with advances in a variety of applications such as content creation, text classification, and question answering (QA). LLMs are trained on a very large amount of text data (and more recently multimodal data), allowing the models not only to recognise patterns and relationships between words and concepts, but also to handle a wide range of tasks, even those for which the models have not been specifically trained and without any explicit supervision \cite{radford2019language,kojima2022large}. LLMs are, in general, more powerful than traditional NLP methods, but they are often considered `black boxes', as the mechanisms behind an LLM decision making are challenging to understand, which makes debugging and bias detection more difficult to investigate. Furthermore, a recent study has tested the performance of ChatGPT in a variety of NLP tasks, and it has been found that the more complex and pragmatic the task (e.g. emotion recognition), the more LLM loses performance \cite{kocon2023chatgpt}. Further, another work found significant biases and inconsistent performance between ChatGPT and GoogleGemini in the detection of sentiment analysis \cite{buscemi2024chatgpt}. However, there are still open possibilities and challenges related to the use of LLMs for automated annotation of data and metadata \cite{tan2024large}. Studies have shown evidence that ChatGPT have outperformed crowd-workers in the text classification of tweets \cite{gilardi2023chatgpt}, and other conventional baselines \cite{chae2023large}. Another work suggests that text classification tasks could be improved with the addition of semantic technologies and knowledge graphs \cite{shi2023chatgraph}. Further, a recent work has shown that GPT models have outperformed the SOTAB open model in Column Property Annotation tasks \cite{korinicolumn}. Based on these findings, our research investigates the performance of ChatGPT (GPT-3.5), GoogleBard, and GoogleGemini in the topic classification task of column headers with a controlled vocabulary of topics. In addition, we compare the classification results between the LLMs and human participants, as well as the effect of topic hierarchy and contextual information.

\subsection{Large Context Window and Retrieval-Augmented Generation}
Large language models often contain outdated or incomplete information, as training data lack real-time updates \cite{he2022rethinking} and domain-specific expertise \cite{li2023chatgpt,shen2023chatgpt}. Furthermore, they are prone to generating irrelevant or factually incorrect content, a phenomenon commonly referred to as \textit{`hallucinations'} \cite{marcus2020next,cao2020factual,raunak2021curious,ji2023survey}. 

Retrieval-Augmented Generation (RAG) systems \cite{lewis2020retrieval} are considered to be a promising solution to these challenges \cite{gao2023retrieval,jiang2023active,guu2020retrieval,borgeaud2022improving}, which combine internal information from LLM with external, and preferably precise, information (e.g. textbook) to improve the accuracy and reliability of information retrieval. RAGs use retrieval systems to index, for example, a textbook stored in a vector database. They have been shown to significantly improve the performance of LLM in a variety of tasks, such as code generation \cite{zhou2022docprompting}, and question-answering (QA) in both an open domain \cite{lewis2020retrieval,peng2023check,izacard2020leveraging,li2022large} and a domain specific setting \cite{cui2023chatlaw}. Although current research shows promising results, there is a lack of understanding of the underlying mechanisms of RAG systems, and recent work has shown limitations in terms of noise and counterfactual robustness, negative rejection, and information integration \cite{chen2024benchmarking}. Moreover, RAGs require a well-functioning retrieval systems which can be more complex to set up.

In comparison, Large Context Window refers to the ability of a language model to process large textual information as input before generating a response. This allows the model to consider the input as context, without external retrieval systems. Following a Large Context Window approach, in this work we leverage knowledge from a controlled vocabulary - the CESSDA topic classification vocabulary - to optimise the topic classification task of column headers. In the following sections, we describe the experimental settings and evaluation metrics used in our investigation.  


\section{Experimental Design and Evaluation}
In this section, we describe the data collection process, experimental design for human and machine topic classification tasks, and evaluation methods. Our experimental design aims to assess the consistency in topic classifications of column headers of three LLMs (ChatGPT using GPT-3.5, GoogleBard and GoogleGemini), comparing them with human-made classifications and investigating the impact of contextual information (e.g., dataset description). All experiments were carried out in February 2024. Initially, the analysis was supposed to include only ChatGPT and Bard. However, Bard was subsequently updated with Gemini, allowing us to also include the latter in the study. We selected OpenAI's ChatGPT and Gemini/Bard from Google because they can be considered the current state-of-the-art (SOTA) LLM, and previous research has also used them for comparison purposes \cite{singh2023chat, rane2024gemini, buscemi2024chatgpt}. A key aspect of our methodology is to provide the same prompt for both human and LLM tasks, allowing evaluation from a neutral point of view. Furthermore, our analysis also considers differences in the classifications based on the hierarchical structure of the topics in the CESSDA controlled vocabulary, which distinguishes between `general' and `specific' topics. For instance, `Education' represents a general topic, while `Higher and Further Education' falls under a more specific subset within the `Education' topic. We aim to investigate how LLMs interpret and classify column headers with respect to both general topics and more specific subtopics, providing insights into the model's comprehension and granularity in the topic classification task.

\subsection{Data Collection}
This work explores how LLMs can be leveraged for the topic classification of column headers using a Controlled Vocabulary (CV). Our analysis uses the Topic Classification CV provided by the Consortium of European Social Science Data (CESSDA)\footnote{\url{https://www.cessda.eu}}. The input column headers were sourced from the CBS Open Data Portal\footnote{\url{https://opendata.cbs.nl/statline/portal}}. We opt for a random dataset selection approach while ensuring diversity between various topics. A total of 10 datasets were selected and we report some of the summary statistics and information below in Table \ref{tab:cbs-dataset}. The chosen datasets varied in the number of columns (ranging from 3 to 68) and in the number of rows (ranging from 340 to 347,130), and were classified under different CBS themes. While we did not include any row-level information in our experiment, we include these statistics here to highlight the diversity among the selected datasets. 

\begin{table}[]
\centering
\caption{The table contains relevant information about the input datasets for the topic classification task.}
\resizebox{\textwidth}{!}{%
\begin{tabular}{lcccc}
\toprule 
\textbf{Title}                                    & \textbf{CBS Identifier} & \textbf{CBS Theme}            & \textbf{N. of Columns} & \textbf{N. of Rows} \\ \hline
Education expenditure and indicators & 80393eng & Education                  & 68 & 280    \\
Health expectancy; since 1981        & 71950eng & Health and Welfare         & 14 & 4536   \\
Listed monuments; region 2023        & 85538eng & Leisure and Culture        & 4  & 347130 \\
Livestock                            & 84952eng & Agriculture                & 3  & 708    \\
Milk supply and dairy production     & 7425eng  & Agriculture                & 11 & 379    \\
Mobility per person, travel modes, travel purpose & 84710eng                & Traffic and Transport         & 12                     & 52800               \\
Plant protection products; sales     & 83566eng & Nature and Environment     & 4  & 494    \\
Population dynamics; month and year  & 83474eng & Population                 & 9  & 380    \\
Social security; key figures         & 37789eng & Labour and Social Security & 19 & 340    \\
Trade and industry; finance, SIC 2008             & 81156eng                & Trade, Hotels and Restaurants & 43                     & 4480                \\ \hline
\end{tabular}%
}
\label{tab:cbs-dataset}
\end{table}

It is important to note that our research is conducted in a zero-shot setting, meaning there is no fine-tuning and pre-training of the models for the topic classification task. Therefore, selecting a vocabulary that aligns with the domain of the datasets becomes crucial to ensure that the topic classification task is effective. In our case, the chosen vocabulary (CESSDA) is relevant to the datasets (from CBS), because both of them are from the social science domain. All input data, CV, code, and results discussed in this work are available on GitHub \footnote{\url{https://github.com/ritamargherita/LLMs-topic-classification}}.

\subsection{LLMs Topic Classification}
\label{sec:summary-statistics}
The prompt used for the LLMs task initiation includes the task specification, input data (i.e. column headers), the CESSDA CV and some formatting constraints. The same prompt was used to query all LLMs, and the same task was executed 10 times for each LLM. For each execution, we refreshed the page and initiated a new chat session. This approach aimed to prevent the risk that prior interactions could influence the execution of the current task and affect the results. 

Additionally, to assess the impact of contextual information, we repeated the process with the inclusion of \verb|*Dataset Description:..| in the prompt inputs. Here, it is important to note that the task with context (i.e., dataset description) could not be performed with GoogleBard due to the size limitation in the allowed prompt. The analysis of the effect of contextual information is therefore performed only with ChatGPT and GoogleGemini. A summary of the prompt is provided below. 

We then labelled the classifications of each column header as follows: \textit{`Specific'} for the CESSDA sub-topics, \textit{`General'} for general topics, \textit{`Other'} distinctively for classification of the CESSDA topic `other', \textit{`Unassigned'} when the classification was not executed, and \textit{`Hallucination'} when the topic classified was not included in CESSDA. It is important here to highlight our deliberate focus on the `Other' topic category. This choice was made because of its potential to indicate that the LLM recognises the absence of a CESSDA topic related to the column header. This behaviour could highlight a nuances understanding of the column header by the LLM, which opts to classify it under the topic of `Other' rather than assigning an unrelated topic. Also, the label `Unassigned' might also indicate that the LLM is not classifying the topic rather than assigning a wrong or random one. However, the former behaviour is preferable, as it suggests an understanding of the topic related to the header is beyond the scope of CESSDA. Also, when the topic is not classified by the LLM, we cannot be sure whether this behaviour is due to the fact that the LLM does not recognise the topic related to the column header within CESSDA, or if it is just an erroneous behaviour. As an evaluation, we performed a Tukey's Honest Significant Difference (HSD) test to assess significant differences between classifications. 

\begin{tcolorbox}[height=0.43\textheight, left=2mm, right=2mm, fontupper=\ttfamily\scriptsize]
Task: Column Header Classification with Controlled Vocabulary \\\\
You are provided with two inputs, below: 1) the column headers of a dataset (in a list format), and 2) a controlled vocabulary of topics (in a CSV format). Your goal is to classify each column header with a relevant topic. The controlled vocabulary has two columns: the 'Topic Label' and 'Topic Description'. For each column header, assign a topic from the controlled vocabulary based on semantic relevance and the definition provided for each topic. The result should be structured in JSON format, where each column header is paired with its assigned topic's label. \\\\
**Constraints: \\
Use only topics provided in the controlled vocabulary, do not add any topics that are not included. \\
Do not change the text of the column headers or topic's label. \\
Only return the output in a JSON format, and no additional text. \\\\
**Inputs: \\\\
*Column Headers (List): \\
$[$h(1), h(2), ....., h(n)$]$ \\\\
*Controlled Vocabulary (CSV Format):\\
Topic Label,Topic Description
\end{tcolorbox}

In addition, we measured the \textbf{Internal Consistency} of each LLM, implementing the Needleman-Wunsch (NW) algorithm \cite{needleman1970general}. The NW algorithm is conventionally used to align genomic sequences, but was adapted here to assess the uniformity of topic classification outcomes in the 10 repeated task executions for each LLM. The pairs of classifications for each LLM and the set of column headers were aligned and scored using the algorithm. We analysed these scores using a multi-way analysis of variance (ANOVA) and post hoc Tukey's HSD tests to investigate differences in outcomes within each LLM. 

Similarly, we measured the similarities in classifications across LLMs - \textbf{Inter-LLMs alignment} - employing again the NW algorithm, to execute and score pairs of classifications across pairs of LLMs. The alignment scores for each pair of LLM were summed and averaged to obtain the \textit{Inter-LLM Alignment} score. ANOVA and Tukey's HSD tests were performed to investigate differences in alignment scores for each pair of LLMs. 

\subsection{Human Topic Classification}
\label{sec:human-topic-classification}
The human topic classifications were performed by three participants: M.M. and T.K., both authors of this paper, and a social scientist who specialises in CBS data. Each participant performed the classification task twice: first without contextual information and then with the dataset descriptions included. Given the minimal difference (less than 5\%) between the topic classifications with and without contextual information, we decided to keep only the classification resulting from the task with context for simplicity.

To measure the agreement between humans and LLM classifications - \textbf{Human-Computer Agreement} - we compute the \textit{joint probability distribution}, which measures the probability of two events happening at the same time. In our case, the events are: 1) each topic classification for a given column header by one LLM, and 2) all human-made classifications for that same column header. In addition, we sum the joint probabilities to measure the probability that the classifications are in agreement between each LLM and the human participants. 

Specifically, equation \ref{eq:probability-1} represents the joint probability $P$ of a human classification $c_{H}$ being the topic $t$ and a machine classification $c_{m}$ also being the topic $t$, given that the human classification belongs to the set of all human classifications $C_{H}$ and the machine classification belongs to the set of all machine classifications $C_{m}$. Subsequently, equation \ref{eq:probability-2}, states that the probability $P$ of human classification $c_{H}$ and machine classification $c_{m}$ being the same topic $t$ is equal to the sum of the joint probabilities of both being $t$. In the following, we report the equations and their variables in Table \ref{tab:joint-probability-variables}.

\begin{equation}
P(c_{H}=t, c_{m}=t | c_{H} \in C_{H}, c_{m} \in C_{m}) = P(c_{H}=t | c_{H} \in C_{H}) \cdot P(c_{m}=t | c_{m} \in C_{m})
\label{eq:probability-1}
\end{equation}

\begin{equation}
P(c_{H}=c_{m} | c_{H} \in C_{H} | c_{m} \in C_{m}) = \sum_{t \in T}P(c_{H}=t , c_{H} \in C_{H}) \cdot P(c_{m}=t | c_{m} \in C_{m})
\label{eq:probability-2}
\end{equation}

\begin{table}[htbp]
    \centering
    \caption{Variable descriptions of the joint probability scoring to measure the agreement between LLMs and humans topic classifications.}
    \begin{tabular}{lll}
        \toprule 
        Variable   & Definition  & \\
        \midrule
        \textbf{Topic}          & $t \in \{1,..,95\}=T$ & A topic $t$ belongs to the set of all topics $T$\\
        \textbf{Classification} & $c \in C$ & A classification $c$ belongs to the set of all classifications $C$\\
        \textbf{Human}          & $h \in \{1,2,3\}=H$ & A human $h$ belongs to the set of all humans $H$\\
        \textbf{Humans classification} &  {$c_{H} \in T$} & A human classification $c_{H}$ belongs to the set of all topics $T$\\
        \textbf{Machine}        & $m \in \{1,2,3\}=M$ & A machine $m$ belongs to the set of all machines $M$ \\
        \textbf{Machine classification} &{$c_{m} \in T$} & A machine classification belongs to the set of all topics $T$\\
        \bottomrule
    \end{tabular}
    \label{tab:joint-probability-variables}
\end{table}


\section{Results}
In this study, we investigated the performance of ChatGPT, GoogleBard, and GoogleGemini in the task of topic classification of column headers, using the CESSDA controlled vocabulary of topics. We also compared human and LLM topic classifications and explored the effect of hierarchical and contextual information on classification outcomes. In the following pages, we report our findings.

\subsubsection{LLMs Topic Classification Summary Statistics} 
Firstly, we have analysed the summary statistics of the LLM topic classifications in both the settings without context and with the context (that is, the description of the dataset) added to the prompt. We used a Tukey HSD test to investigate any significant differences in classified topics, based on the labels introduced in \ref{sec:summary-statistics}. To reiterate, the labels are 1) \textit{`Specific'} for the CESSDA sub-topics, 2) \textit{`General'} for general topics, 3) \textit{`Other'} for the classification of the exact CESSDA topic `Other', 4) \textit{`Unassigned'} when the classification was not executed, and 5) \textit{`Hallucination'} when the topic classified was not included in CESSDA. 

For the task \textbf{without context} - i.e. without adding the description of the data set in the prompt - Tukey's HSD test revealed significant differences between the \textit{ChatGPT-GoogleBard} pair in all topic classifications except for \textit{`Specific'} topics ($p(2)=0.02, p(3)=0.01, p(4)=0.03, p(5)=0.03$). The \textit{ChatGPT-GoogleGemini} pair, instead, showed significant differences only in the \textit{`Other'} topic classifications ($p(3)=0.01$). Lastly, the \textit{GoogleGemini-GoogleBard} pair had significant differences for \textit{`Unassigned'} ($p(4)=0.02$) and \textit{`Hallucinated'} ($p(5)=0.01$) topic classifications. For the task \textbf{with context} - with the dataset description added to the prompt -, we found a weak significant difference in the classification of topics between the ChatGPT-GoogleGemini pair for the \textit{`General'} topics ($p(2)=0.0469$), and a stronger difference for the \textit{`Other'} topics ($p(3)=0.01$). 

\begin{figure}[htbp]
    \centering
    \includegraphics[width=1\linewidth]{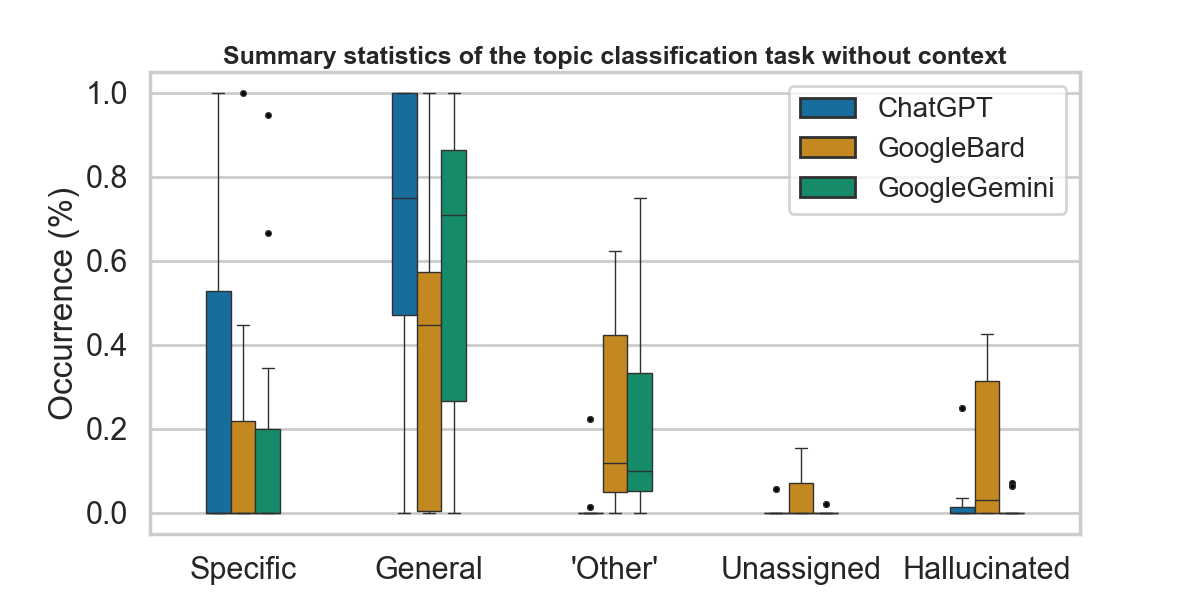}
    \caption{Summary of the topic classification task by the three LLMs, in the setting with no contextual information added to the prompt. We show the distribution of the topics classified based on 5 labels: `Specific' topics, `General' topics, the `Other' topic, `Unassigned' topics and `Hallucinated' topics, i.e. outside of the controlled vocabulary.}
    \label{fig:summary-boxplot-no-context}
\end{figure}

\begin{figure}[htbp]
    \centering
    \includegraphics[width=1\linewidth]{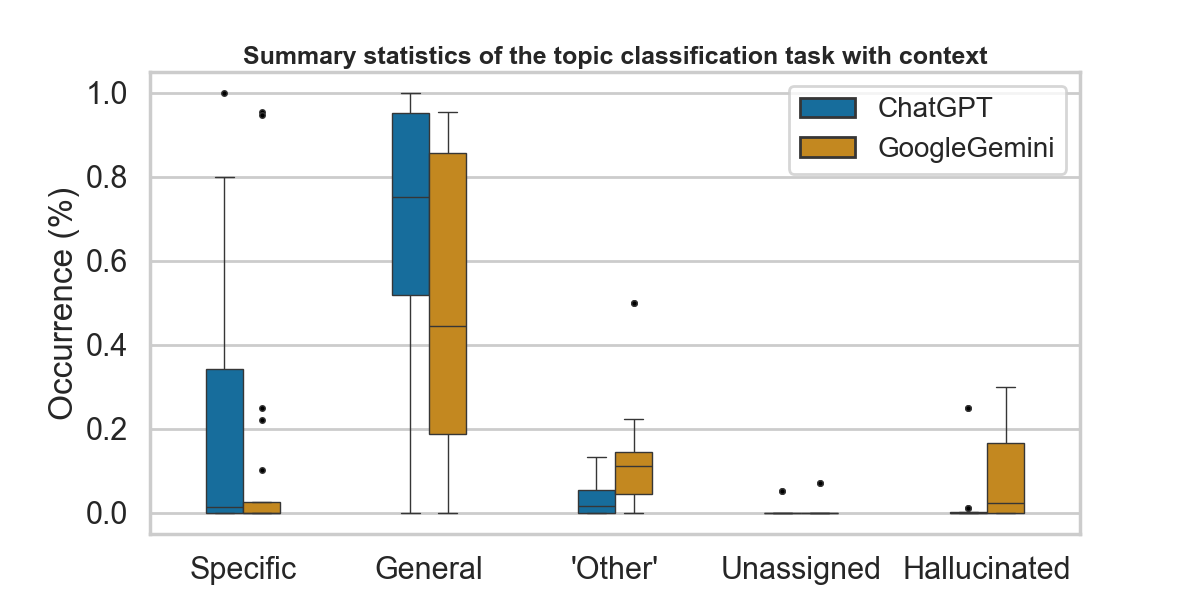}
    \caption{Summary of the topic classification task by the three LLMs, in the setting with contextual information added to the prompt. We show the distribution of the topics classified based on 5 labels: `Specific' topics, `General' topics, the `Other' topic, `Unassigned' topics and `Hallucinated' topics, i.e. outside of the controlled vocabulary.}
    \label{fig:summary-boxplot-context}
\end{figure}

We present two boxplots below, where Figure \ref{fig:summary-boxplot-no-context} reports the distribution of classified topics in the setting without context, and Figure \ref{fig:summary-boxplot-context} when the task was performed with the addition of context. We can see that in \ref{fig:summary-boxplot-no-context} GoogleBard showed fewer instances of the classification of `Specific' and `General' topics. It also assigns the `Other' topic more frequently, particularly compared to ChatGPT. Moreover, instances of `Hallucinated' topics were more prevalent compared to the two other LLMs. In \ref{fig:summary-boxplot-context}, instead, we report the distribution of classified topics based on the labels only for ChatGPT and GoogleGemini, because we were unable to perform this task with GoogleBard, as previously mentioned. This boxplot indicates that ChatGPT and GoogleGemini had, in general, a similar distribution of classified topics.

In summary, our findings and the Tukey's results suggest that ChatGPT and GoogleGemini are more similar in performances compared to GoogleBard in the classification task. They also show that GoogleBard is more likely to assign the topic `Other', indicating a tendency to abstain from making a definite classification. In addition, the results suggest that contextual information does not have a strong impact on the types of classified topics.

\subsubsection{LLM Internal Consistency}
With this measure, we assessed the internal consistency of the LLM in the classification task of each dataset in the 10 task executions. Using the NW algorithm, we measured the internal consistency, and we evaluated it using Multi-Way ANOVA and Tukey's HSD tests. For the task \textbf{without context}, the overall consistency scores for each LLM were: $ChatGPT = 0.52$, $GoogleBard = 0.11$ and $GoogleGemini = 0.81$, where 1 is absolute consistency for all several executions for each dataset. Multi-way ANOVA showed a significant effect on consistency scores based on the dataset (p = $1.87^{-7}$). Tukey's test confirmed significant differences in consistency between: the ChatGPT-GoogleBard pair($p=0.007$), and GoogleGemini-GoogleBard pair ($p=0.0001$). No differences were found between the ChatGPT-GoogleGemini pair ($p=0.3$), suggesting similar consistency scores across all task executions and datasets. For the task \textbf{with context}, we found the overall internal consistency scores of $ChatGPT = 0.46$ and $GoogleGemini = 0.51$. ANOVA and Tukey's test did not find significant differences in consistency scores between these LLMs, supporting the above findings. 

These results indicate that, in general, GoogleGemini appears as the LLM that is more consistent in the classification of topics across repeated task executions. GoogleBard, instead, shows much lower scores for the internal consistency measure. The ANOVA test also suggests that the dataset in which the column headers are classified can have an impact on the consistency score. This result needs further investigation, to evaluate whether there is a correlation between different aspects of the datasets (e.g. number of columns, domain, expressivity of column headers) and the internal consistency score. Furthermore, it appears that GoogleGemini and ChatGPT had similar internal consistency scores across all datasets, and no significant differences were found between these two LLMs even when the task was performed in context. 

\subsubsection{Inter-LLMs Alignment}
To measure the agreement of the topic classifications between LLMs, we calculated the Inter-LLMs Alignment score using the NW algorithm and performed ANOVA and Tukey's tests. We computed the alignment scores for each LLM pair in both tasks with and without context. In all cases, the alignment scores were approximately 0, suggesting different classified topics for each LLM. For the task \textbf{without context}, ANOVA revealed that the datasets to which the column headers belonged had significant effects for the ChatGPT-GoogleBard and GoogleGemini-GoogleBard pairs ($p=1.05^{-8}$ and $p=2.75^{-9}$ respectively). No significant effects from the dataset were found for the GoogleGemini-ChatGPT pair. Furthermore, for the task \textbf{with context}, ANOVA found no significant effect of the dataset between the ChatGPT-GoogleGemini pair, supporting previous findings. Similarly to the results from the Internal Consistency scoring, the effect that the dataset might have on LLM performance needs further investigation. However, no significant effect was found for the ChatGPT-GoogleGemini pair for both the task with and without context, indicating that these two LLMs might have comparable underlying processes and performances. 

\subsubsection{Human-Computer Agreement}
We calculated the Human-Computer Agreement (HCA) scores based on the joint probability metrics introduced in \ref{sec:human-topic-classification}. The scores are reported in Table \ref{tab:hca-scores}, where an agreement score of 1 indicates agreement between LLM classification and at least one human classification. The table reports the scores for the tasks with and without context, as well as based on the hierarchy of topics in CESSDA. In the table, the `Exact Match' score represents the agreement between the human and machine classification when the topics are exactly the same. The `Close Match' score, instead, involves mapping the topics to their general topic in the CESSDA controlled vocabulary. In other words, while `Exact Match' requires exact agreement between topics (e.g. both the human and machine classifications are the CESSDA topic of \textit{Education}), `Close Match' allows for slight variations, as long as both machine and human classified topics belong to the same overarching CESSDA term (e.g. \textit{Education} and \textit{Higher and Further Education}). We find that ChatGPT classifications aligns most closely with the human-made classifications in the `Close Match' setting for both tasks with and without context, and it also shows a slightly higher HCA for the task with context compared to the one without. Interestingly, GoogleGemini shows lower HCAs for the task with context compared to the one without context. Although we lack sufficient data for statistical significance between HCAs for tasks with and without context, these initial findings again support that contextual information may not have a significant effect on the classification task.

\begin{table}[]
    \centering
    \caption{The table shows for each LLM and settings (context and no-context) the agreement between machine and human classifications, where 1 is complete agreement and 0 is no agreement at all.}
    \begin{tabular}{lccccl}\toprule
        \multicolumn{1}{c}{} & \multicolumn{2}{c}{No Context} & \multicolumn{2}{c}{With Context}
        \\\cmidrule(lr){2-3}\cmidrule(lr){4-5}
                        & Exact Match  & Close Match  & Exact Match & Close Match\\\midrule
        ChatGPT         & 0.29      & 0.5   & 0.33   & 0.53\\
        GoogleGemini    & 0.28      & 0.46  & 0.15   & 0.37\\
        GoogleBard      & 0.24      & 0.31  & X      & X\\\bottomrule
    \end{tabular}
    \label{tab:hca-scores}
\end{table}


\section{Conclusion and Future Work}
In this work we propose a novel approach that leverages LLMs for text classification of column headers with a topic controlled vocabulary. Our experimental design focuses on exploring the impact of contextual and hierarchical information in the topic classification task. We have evaluated the performance of three LLMs (ChatGPT, GoogleBard, and GoogleGemini) through various metrics: 1) we investigated the nature of classified topics, including the hierarchical structure of the controlled vocabulary; 2) we measured the internal consistency of each LLM in the classification task; 3) we evaluated the alignment of classified topics across LLMs; and 4) we measured the classification agreement between the LLMs and human participants. Our findings suggest that, in general, ChatGPT and GoogleGemini outperform GoogleBard in the column header topic classification task. Interestingly, contextual information appears to have no significant effect on the consistency and agreement of the classification tasks for the LLMs. We also did not find strong evidence suggesting that hierarchical information affects the classification task. Moving forward, our goal is to perform this investigation with a larger corpus of input data to better support statistical analysis and explore whether LLMs can capture semantic similarities based on the relationships between columns. We also intend to further explore the RAG approach by using available tools (e.g. the OpenAI RAG plug-in) in order to incorporate larger controlled vocabularies, as well as engage with additional domain experts to better validate and refine the results of the classification task. Although this study represents an initial exploration, it serves as a starting for the topic classification task of column headers, and it provides a groundwork approach for enhancing metadata with column-level information and advance the Findability, Accessibility, Interoperability, and Reusability of datasets on the Web.

\subsection{Acknowledgement}
The authors thank Emma Beauxis-Aussalet for her help with the mathematical notation of the Human-Machine Agreement scoring function. In addition, we acknowledge that ChatGPT was used to generate and debug part of the Python and \LaTeX code used in this work. This work is funded by the Netherlands Organisation of Scientific Research (NWO), ODISSEI Roadmap project: 184.035.014.

\bibliographystyle{vancouver.bst}
\bibliography{mybibfile}

\end{document}